# Data Partitioning for Parallel Entity Matching

Toralf Kirsten, Lars Kolb, Michael Hartung, Anika Groß, Hanna Köpcke, Erhard Rahm

Department of Computer Science, University of Leipzig

04109 Leipzig, Germany

{tkirsten,kolb,hartung,gross,koepcke,rahm}@informatik.uni-leipzig.de

## ABSTRACT

Entity matching is an important and difficult step for integrating web data. To reduce the typically high execution time for matching we investigate how we can perform entity matching in parallel on a distributed infrastructure. We propose different strategies to partition the input data and generate multiple match tasks that can be independently executed. One of our strategies supports both, blocking to reduce the search space for matching and parallel matching to improve efficiency. Special attention is given to the number and size of data partitions as they impact the overall communication overhead and memory requirements of individual match tasks. We have developed a service-based distributed infrastructure for the parallel execution of match workflows. We evaluate our approach in detail for different match strategies for matching real-world product data of different web shops. We also consider caching of input entities and affinity-based scheduling of match tasks.

## 1. INTRODUCTION

Entity matching or entity resolution is the process of identifying entities (i.e., data instances, records) referring to the same real world object. This task is of critical importance for data integration and especially challenging for web data. Hence, numerous approaches, frameworks and tools for entity matching have been proposed and developed in the recent past [1, 6, 11, 14]. Typically, the similarity between entities is determined by applying several matchers (e.g., comparing attribute values with some string similarity measure) and combining the individual similarity values to derive a match decision for a pair of entities. For example, two product entities may be assumed to match if both their product titles and product descriptions are very similar. In some systems, the combination of matcher similarities is determined by training-based machine learning approaches such as decision trees [13].

The execution of such match strategies poses typically high resource (CPU, memory) demands. A straightforward approach evaluates the different matchers on the Cartesian product of input entities, i.e., it implies a quadratic complexity of $O(n^2)$ for finding matches in *n* input entities. Such an approach obviously has severe scalability restrictions for larger sets of entities and is rarely affordable for online data integration, e.g., within data mashups. A recent performance evaluation of learning and non-learning entity resolution approaches [12] revealed substantial efficiency problems of current techniques. For matching subsets of the bibliographic datasets DBLP and Google Scholar (2,600 vs. 64,000 objects) evaluating the Cartesian product took up to 75 h for a single attribute matcher. The execution times increase even more if matching on multiple attributes is applied.

*Blocking* and *parallel matching* are two options to improve the efficiency of entity matching. Blocking avoids the evaluation of the complete Cartesian product of entities but reduces the search space for entity matching [2]. Typically this is achieved by some kind of semantic partitioning by grouping similar entities within clusters (or "blocks") and by restricting entity matching to entities from the same block. For example, one could partition products by manufacturer and compare only products with the same manufacturer during entity matching.

Parallel matching aims at improving performance by splitting a match computation into several match tasks and executing these tasks in parallel, e.g., on multi-core servers or on a cloud infrastructure. Surprisingly such a parallel matching has received little attention so far and will be studied in this paper. A key prerequisite for effective parallel matching is the suitable partitioning of the input entities which influences the processor utilization, communication overhead and load balancing. Furthermore, entity matching is typically memory-sensitive so that the definition of match tasks should consider the available memory that is typically shared among several cores. Furthermore, it is non-trivial to effectively combine blocking and parallelization since blocking may result in blocks of largely different size.

Our contributions are as follows:

- We propose two general partitioning strategies for generating entity match tasks that can be executed in parallel. The first approach is based on the Cartesian product of entities while the second, more sophisticated strategy applies a combination of blocking and parallelization. Both partitioning approaches aim at avoiding memory bottlenecks and load imbalances for the resulting match tasks. The partitioning schemes are applicable to different match strategies, e.g. combining several matchers.
- We present a flexible, service-based infrastructure for parallel entity matching on different hardware configurations. Caching and affinity-based scheduling of match tasks are supported to reduce communication requirements.
- We implemented the proposed partitioning strategies and infrastructure and perform a comprehensive evaluation for matching real-world product data. The results show the effectiveness and scalability of the proposed approaches for different match strategies.

In the next section, we introduce some preliminaries on entity matching and the assumed computing environment. Section 3 describes the two strategies for partitioning the input data. Section 4 outlines the match infrastructure of our approach. Section 5 describes the performed experiments and evaluation. We discuss related work in Section 6 and conclude with a summary and outlook.

## 2. PRELIMINARIES

The input of entity matching consists of a set of entities. Entities are described by attributes, such as product name, description, manufacturer or product type. In this paper we focus on the common case where all entities to be matched reside already in a single dataset. The approaches can be extended to match entities from different sources as we will briefly discuss in Section 3.3. Entity matching determines a set of correspondences ($e_1$, $e_2$, $sim$) indicating the equivalence similarity (from the interval [0,1]) for pairs of entities $e_1$ and $e_2$. All entity pairs with a similarity exceeding a specific threshold are assumed to match.

We assume that match results are determined according to some entity matching workflow as supported by our framework FEVER [12]. Entity matching workflows either evaluate the Cartesian product of input entities or first apply a blocking operator to reduce the search space for matching. Blocking results in a set of blocks / clusters of related entities and, ideally, matching can be restricted to comparing entities of the same block (for an exception see Section 3.2).

Entity matching itself is specified by a *matching strategy* entailing the execution of one or several matchers, e.g., the similarity computation for a specific attribute and similarity measure. There are many possibilities to combine the results of several matchers. We assume one of the approaches of FEVER, i.e., the combination is either specified manually (e.g., intersection of matches) or determined by a training-based model derived by a machine learning method such as SVM, decision tree or logistic regression. In this paper we largely treat the match strategies as black boxes. Our approach for parallel matching focuses on the partitioning the input data and defining multiple match tasks on input partitions that can be executed in parallel. This kind of data parallelism is very generic and can be utilized for different match strategies. In our evaluation, we will consider two match strategies with several matchers.

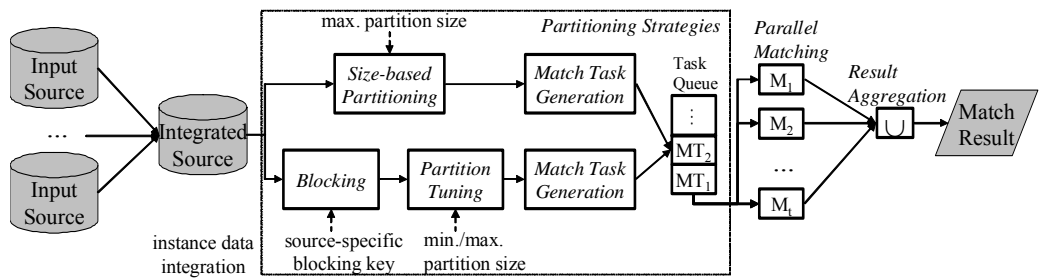

**Figure 1: Match Workflows with Partitioning Strategies and Parallel Matching**

The parallel match execution uses a given computing environment *CE* = (*#nodes*, *#cores*, *max_mem*) consisting of *#nodes* loosely coupled computing nodes each providing *#cores* cores (processors) and a maximum main memory *max_mem* to be shared by the cores of a node. For simplicity we assume homogeneous nodes with the same number of cores and memory size. However, the model can easily be extended for heterogeneous configurations. Furthermore, we assume that each node can equally access the input data, e.g., from a shared server (shared disk) or by replicating the input data among the nodes. This data placement enables a high flexibility for scheduling match tasks and thus for dynamic load balancing. As we will discuss in Section 4, input data may also be cached in main memory to improve performance.

## 3. PARTITIONING STRATEGIES

To enable parallel entity matching on multiple nodes and cores, we first apply a partitioning strategy to partition the input data source and to generate multiple match tasks that can be independently executed. As shown in Figure 1, we distinguish two different partitioning strategies: (1) *size-based partitioning* to evaluate the Cartesian product for entity matching and (2) *blocking-based partitioning* to combine blocking with parallel matching. Strategy (2) consists of a separate phase called *partition tuning* to deal with large block size differences. Both partitioning approaches end in a step to generate match tasks that compare entities of the input partitions. In our infrastructure (Section 4) the match tasks are maintained in a central task list and are executed in parallel on the computing nodes. The union of the individually determined match results finally gives the complete result.

In the following we describe the two partitioning approaches in more detail for the case of a single input dataset. Finally, we briefly discuss how we can deal with several input datasets to be matched.

### 3.1 Size-based Partitioning

Size-based partitioning is the simplest approach and is applied for evaluating the complete Cartesian product of input entities. The goal is to split the input entities into equally-sized partitions so that each match task has to match two such partitions with each other. This approach is thus very simple and the equally sized partitions promise a good load balancing and scalability to many nodes.

The main consideration is to choose a suitable partition size *m* which also determines the number of partitions *p* ($p=\lceil n/m \rceil$ for *n* input entities) and the number of match tasks (see below). The number of partitions and match tasks should clearly be higher than the number of available cores in the system. On the other hand, the partition size should not be too small since otherwise the overhead for communication and starting/terminating match tasks would be comparatively high compared to the actual match work.

The partition size also influences the memory requirements of a match task and is thus restricted by the available memory. This is because entity matching typically takes place in main memory so that evaluating the

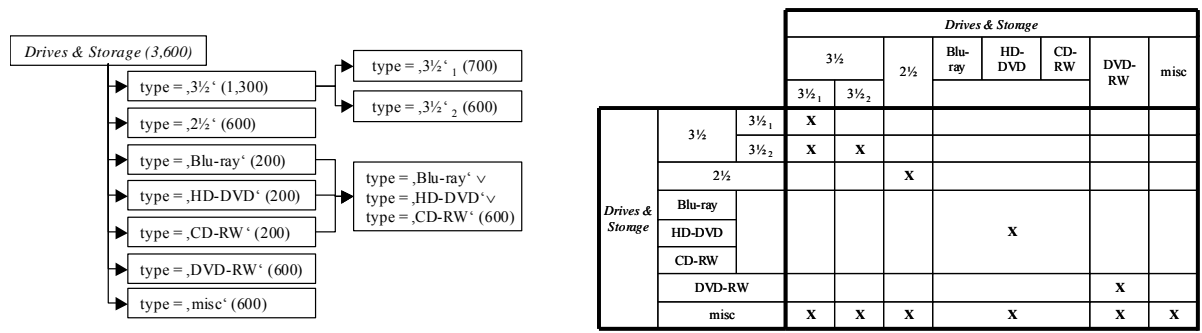

**Figure 3: Blocking and partition tuning example (left) and corresponding match task generation (right)**

|       | Input Set |       |     |           |       |
|-------|-----------|-------|-----|-----------|-------|
|       | $P_1$     | $P_2$ | ... | $P_{p-1}$ | $P_p$ |
| $P_1$ | X         |       |     |           |       |
| $P_2$ | X         | X     |     |           |       |
| ...   | ...       | ...   | ... |           |       |
| $P_{p-1}$ | X     | X     | ... | X         |       |
| $P_p$ | X         | X     | ... | X         | X     |

**Figure 2: Match task generation for size-based partitioning**

Cartesian product of two partitions of size $m$ leads to memory requirements of $O(m^2)$. The exact memory requirements of a match task depend on the chosen match strategy and implementation, e.g., which matchers are executed, whether or not a machine learning model is applied, and the number of intermediate correspondences. Assuming an average memory requirement of $c_{ms}$ per entity pair for match strategy $ms$ (for the correspondences/similarity values etc.), the memory requirement per match task is about $c_{ms} \cdot m^2$. These memory requirements have to be compared with the available memory per core (or per parallel thread) which can be approximated by $max\_mem / \#cores$ for our model of the computing environment. The memory-restricted partition size $m$ can thus be estimated by $m \leq \sqrt{max\_mem/(\#cores \cdot c_{ms})}$.

A memory-efficient match strategy may only consume about 20 B per entity pair. For $max\_mem=2$ GB and $\#cores=4$ we could use 500 MB per match task and could thus match partitions of maximal size $m = \sqrt{500MB/20B} = 5,000$ entities. By contrast, a memory-consuming match strategy, such as typical for learner-based approaches, might require $c_{ms} = 1$kB. In this case the maximum partition size would decrease to only $m = \sqrt{500MB/1kB} \approx 700$ entities. In our evaluation, we will experimentally determine the impact of different partition sizes.

The generation of match tasks is straight-forward for size-based partitioning. To evaluate the Cartesian product the entities of each partition need to be compared with all entities of the same partition as well as with the entities of all other partitions. Figure 2 illustrates the resulting match tasks for $p$ partitions ($P_1, ..., P_p$); the cross signs indicate the two input partitions per match task. For $p$ partitions, size-based partitioning generates $p+(p(p-1)/2)$ match tasks, i.e., we match two partitions $P_i$ and $P_j$ if $i \leq j$.

## 3.2 Blocking-based Partitioning

Evaluating the Cartesian product does not scale and is thus only feasible for small-sized match problems. For larger match problems the use of blocking becomes necessary to reduce the scope for matching. Our second partitioning strategy deals with the results of a blocking step and uses a partition tuning step before the generation of match tasks.

Blocking entails a logical partitioning of the input entities such that all matching entities should be assigned to the same output partition, also called cluster or block. By restricting the match comparisons to the entities of the same block the match overhead can often drastically be reduced. Our partitioning strategy is largely independent of the particular blocking approach chosen, e.g., Canopy Clustering [15] or Sorted Neighborhood [9]. The partitioning criterion is application-dependent and should enable matching entities assigned to the same block. In the simplest case, one may use a range partitioning on specific attributes for blocking, e.g., to partition products by product type or manufacturer, or to partition publications by publication year or publication venue.

Unfortunately, due to missing data values or other data quality issues in real-world data (e.g., products without product type information) it may not always be possible to assign entities to a unique block. We therefore assign such entities to a dedicated *Miscellaneous* (*misc*) block. Entities of this block have to be matched against the entities of all blocks.

The output blocks may largely differ in their size depending on the entity values and the applied blocking approach. Hence, simply using one match task per block would mostly not result in an effective parallel matching but poor load balancing (skew effects) or/and high communication overhead. On one hand, large blocks could dominate the execution time for matching and require more memory than available. On the other hand, very small blocks would result in tiny match tasks with a relatively high communication and scheduling overhead.

To deal with these problems we therefore perform a *partition tuning* to split or combine blocks:

- First, we determine all large blocks for which the memory requirements exceed the maximal size according to the estimation described in Section 3.1. These blocks are then split into several equally sized partitions obeying the size restriction. Match task generation has to consider that all these sub partitions have to be matched with each other.
- Second, we aggregate all smaller blocks (e.g., with a size below some fraction of the maximal partition size) into larger ones. The reduction in the number of partitions results in fewer match tasks and thus a reduced communication and scheduling overhead. On the other hand, by comparing all entities of the aggregated blocks with each other we may introduce some unnecessary comparisons compared to the original blocking. In our evaluation we will study this performance tradeoff.

For *match task generation* we need to distinguish three different cases:

- All normal (non-*misc*) blocks that have not been split and aggregated result in one match task that matches the entities of the block with each other.
- Blocks that have been split into $k$ sub-partitions result into $k+(k\cdot(k-1)/2)$ match tasks to match these sub-partitions with each other.
- The *misc* block (or its sub-partitions in case it had to be split) has to be matched with all (sub-) partitions.

The example in Figure 3 (left) illustrates blocking and partition tuning for a small set of 3,600 Drives & Storage products. We first block the products by their product type. The resulting blocks vary in their size from 200 to 1,300. The *misc* block is assumed to hold 600 products with unknown product type. For a maximal (minimal) partition size of 700 (210) entities, partition tuning would split the largest partition ('3½' drives) into two partitions and aggregate the smallest blocks 'Blu-ray', 'HD-DVD' and 'CD-RW' into one partition of size 600. The resulting match tasks are shown in Figure 3 (right). For well-sized "non-misc" blocks we only compare the entities within this partition within one match task (e.g., product types '2½' and 'DVD-RW'). Similarly, we match the aggregated partition within one match task. The two sub-partitions of the split block of '3½' entities result in three match tasks. Finally, the *misc* partition is matched against all other partitions resulting in six match tasks. Hence, we generate a total of 12 match tasks for the example. A size-based partitioning would have created six partitions and 6+6·5/2=21 match tasks.

### 3.3 Matching Multiple Input Sources

The proposed approach for partitioning and entity matching can also be applied when data from two or more input sources should be matched. A straight-forward approach to deal with this situation is to first take the union of the input sources and apply the proposed approaches for partitioning and parallel matching on the combined result (as indicated in Fig. 1). In the case of heterogeneous input schemas we first have to align (match) differently named but equivalent attributes, either manually or with the help of a schema matching tool. In more complex cases we may also have to apply some data transformations (e.g., to split complex address attribute into several simpler attributes) to ensure the comparability of match attributes in the combined data source.

In the special case when the input sources are duplicate-free we can utilize this for a reduced match effort by not matching entities of the same source with each other. For example, for evaluation the Cartesian product we would partition both sources into equally sized partitions and match each of the n partitions of the first with each of the m partitions of the second source ($m*n$ match tasks compared to $(m+n)*(m+n-1)/2$ tasks for a single, combined source). We can also apply blocking-based partitioning for two inputs by applying the same blocking approach to both sources and only match corresponding blocks with each other; entities in misc blocks need to be matched with all other blocks of the other source. For tuning the block sizes we can employ the same approaches as discussed in the previous subsection.

## 4. MATCH INFRASTRUCTURE

We implemented a distributed service-based infrastructure to execute entity matching workflows in parallel according to the proposed strategies. Figure 4 shows the architecture of the system. It consists of different kinds of services, namely a *workflow service*, a *data service*, and multiple *match services,* running on several loosely coupled nodes (servers or workstations). The workflow service is the central access point of the infrastructure managing the execution of match workflows. Since we currently restrict parallel processing on matching, we use the workflow service to perform the pre- and post-processing including blocking and partitioning of the input data as well as merging the final match result. The workflow service initially generates all match tasks and maintains them in a common task list for scheduling them among the match services. The input data, blocking output and match results are stored and managed by a *data service*. We currently use a central DBMS server for this purpose. All services are implemented in Java and use the Remote Method Invocation (RMI) protocol for communication.

Match services run on dedicated nodes and can concurrently execute several match tasks within threads (one match task per thread). The number of threads may be set to the number of cores but can also be chosen differently. Each match service can utilize a *partition cache* to temporarily store entity partitions. Cached partitions can be used by all threads of a match service to reduce the amount of communication for repeatedly fetching the same data from the data service. The cache size per match service is configured by the maximal number of cached partitions $c$; $c=0$ means that caching is disabled. The caches are managed according to a LRU replacement strategy. That is when a newly fetched partition is to be stored in a full cache, it replaces the partition with the oldest access time.

The workflow service is responsible for assigning unprocessed match tasks in its task list to the match threads for execution. When partition caching is enabled, we use a simple strategy to perform an *affinity-based scheduling* aiming at assigning match tasks to match services where needed partitions are already cached. To maintain an approximate cache status at the workflow service, match services report which partitions they cache whenever a match task is completed. By sending the cache information together with the match results of a match task the cache status at the workflow service is maintained without additional messages. After the completion of a match task is reported, the workflow service assigns a new match task to the respective (preferably a task for which needed partitions are cached at the thread's match service). This task selection and assignment is repeated as long as there are still open tasks to execute. This simple scheme supports a distributed match processing with dynamic load balancing on the one hand and locality of match processing to exploit caching on the other hand. Furthermore, the scheduling approach can easily cope with heterogeneous

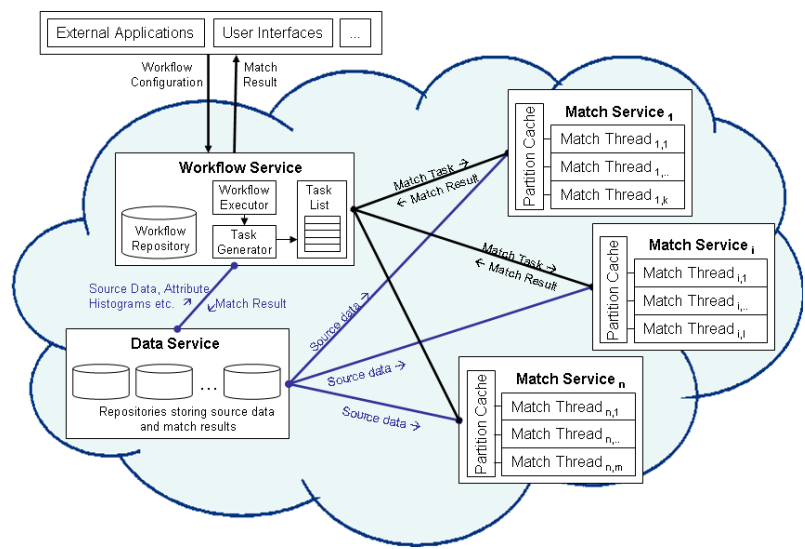

**Figure 4: Service-based infrastructure for parallel entity matching**

nodes (different speeds, different number of cores/threads).

New match services can be added on demand by starting the new service on a dedicated physical machine and informing the workflow service about the new match service. Similarly, match services can be taken out from match processing. This makes the infrastructure dynamic and adaptable, i.e., depending on the complexity of the match problem we can adjust the number of used match services. The architecture is also robust against failures of match services. If a match service does not respond anymore, the workflow service can assign the match tasks of the respective service among the remaining services.

## 5. EXPERIMENTS

We conducted a series of experiments to evaluate the effectiveness of the proposed partitioning strategies for parallel entity matching for different hardware configurations, match problems and match workflows. After a description of the experimental setup we first focus on parallel matching using several cores of a single server. We then analyze the scalability of parallel matching for multiple nodes. Finally, we evaluate the impact of caching on execution times.

### 5.1 Experimental setup

We ran our experiments on up to six nodes with four cores. Each node has an Intel(R) Xeon(R) W3520 4x2.66GHz CPU, 4GB memory, and runs a 64-bit Debian GNU/Linux OS with a 64-bit JVM. Both the workflow service and the data service run on a dedicated server so that up to four nodes (16 cores) are used for parallel match processing. We use 3 GB main memory (heap size) per node for matching. Following the notation of Section 2.1, the setup can be described with $CE = (4, 4, 3GB)$. In one experiment, we vary the number of threads from 1 to 8. We initially disable partition caching ($c=0$) but study the effect of caching in Section 5.4.

The main input dataset for entity matching contains about 114,000 electronic product offers from a price comparison portal. Each product is described by 23 attributes. In addition to this large scale match problem we also use a representative subset of 20,000 products to define a smaller-scale match task.

The first match strategy (WAM) executes two matchers (edit distance on product title, TriGram similarity on product description) and calculates a weighted average of the two matcher results. The second match strategy (LRM) executes three matchers (Jaccard, TriGram, and Cosine attribute similarity) and uses the machine learning approach Logistic Regression to combine their results. The WAM strategy applies an internal optimization to reduce memory requirements by eliminating all correspondences of a matcher with a similarity too low for reaching the combined similarity threshold. For example, if matching entity pairs need to have an average similarity of at least 0.75 for two matchers, all correspondences with a single-matcher similarity below 0.5 can be discarded since the average similarity threshold cannot be reached anymore.

In this study we do not evaluate matching effectiveness but focus on efficiency, in particular execution times for matching. In [12,13] we have evaluated the effectiveness of different match strategies for a similar match problem on product entities and found learner-based approaches to be generally more effective than simple matcher combinations such as WAM when several matchers on different attributes need to be combined .

### 5.2 Parallel matching on a single node

We first analyze the impact of the number of treads as well as of the maximal and minimal partition size for the small match problem on one node.

For a 4-core node we consider between 1 and 8 threads for matching on one node. We evaluate the Cartesian product (size-based partitioning) for the small match problem using a conservative partition size of 500 entities. Figure 5 shows the resulting execution time and speedup for both match strategies WAM and LRM. We observe that our size-based partitioning enables an effective parallel matching resulting in improved execution times. For one and two threads the two match strategies perform similarly but for more threads the memory-optimized WAM strategy outperforms LRM because more threads reduce the amount of memory per match task. WAM also benefits more from parallel matching with an almost linear speedup for up to 4 threads (factor 3.5) while LRM achieves a speedup of up to 2.5. WAM can benefit only marginally from more than 4 threads and LRM not at all. Hence we will use at most 4 threads (= #cores) per node in our further experiments.

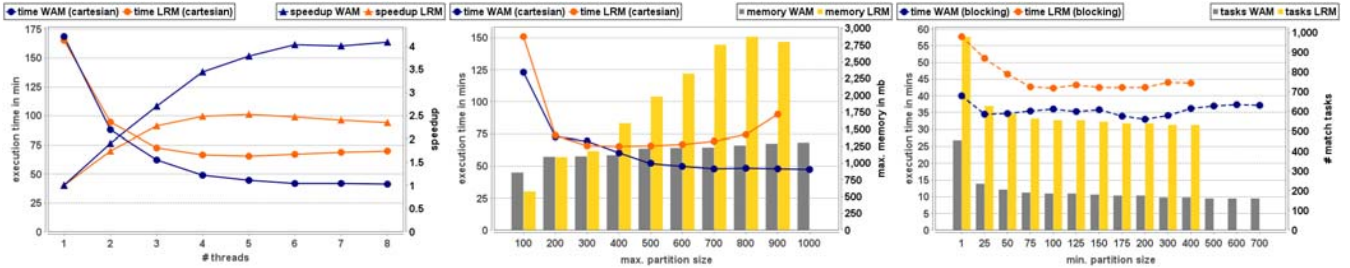

**Figure 5: Speedup per multi-processor node**

**Figure 6: Influence of the maximum partition size**

**Figure 7: Influence of the minimum partition size**

Next we evaluate the influence of the (maximum) partition size for evaluating the Cartesian product with *size-based partitioning* introduced in Section 3.1. Figure 6 shows the execution time as well as the consumed memory for partition sizes between 100 and 1000 entities (using 1 node running 4 match threads). We observe that increasing the partition size from 100 to 200 strongly improves execution times especially for LRM (factor 2). This is mainly because of the strong reduction of the number of match tasks and thus for communication and scheduling overhead. Furthermore, larger match tasks better utilize the available processing capacity. Further increasing the partition size leads to additional execution time improvement for WAM. For LRM, on the other hand, the memory consumption (and the amount of paging I/O) grows with larger partitions and execution times start to deteriorate for more than 500 entities per partition. For our further experiments we use the favorable maximal partition sizes of 500 for LRM and 1,000 for WAM.

As pointed out in Section 3.2, we also need to consider a minimal partition size in the case of *blocking-based partitioning*. We analyzed the influence of this parameter for blocking on the manufacturer attribute. Again the evaluation was done for the smaller match problem on a single node executing 4 parallel threads (partition size=1000/500). Figure 7 shows the execution times for the two match strategies for minimal partition sizes between 1 (i.e., no merging of small partitions) and 700, as well as the number of match tasks. We observe that merging small partitions is especially effective for combining the smallest blocks as it significantly reduces the number of match tasks and the associated overhead. These improvements are achieved for both match strategies but are especially pronounced for LRM. LRM suffers from a much higher number of match tasks than WAM (and thus a higher execution time) because of the smaller maximal partition size. For the remaining experiments we choose beneficial minimum partition sizes, namely 200 for WAM and 100 for LRM.

### 5.3 Parallel matching on multiple nodes

To analyze the scalability of our partitioning strategies and infrastructure we now evaluate parallel matching for up to 4 nodes and 16 cores (threads). Particularly, we perform experiments for the small as well as for the large match problem applying the partition sizes determined in the previous experiments.

Figure 8 shows execution times and speedup results for up to 16 cores for both size-based partitioning (continuous lines) and blocking-based partitioning (dashed lines) with the two match strategies on the small-scale problem. The configurations up to 4 threads refer to the use of 1 node, for up to 8 threads to 2 nodes and so on. We observe that execution times scale linearly for up to 16 cores for both partitioning strategies and both match strategies. As expected, the use of blocking improves execution time. Our blocking-based partitioning with its partition tuning proves to be very effective since we achieve the same high speedup values (of up to 14 for 16 cores) than for the simpler size-based partitioning on the Cartesian product. LRM is consistently less efficient than WAM due to its increased memory consumption and the larger number of match tasks.

Figure 9 shows the performance of blocking-based partitioning for matching the full dataset of 114,000 entities. The evaluation of the Cartesian product (ca. 6.5 billion entity pairs) was too time-consuming and not further regarded. Blocking-based partitioning resulted in about 1,200 match tasks for WAM compared to 3,900 tasks for LRM due to its more limited partition sizes. More than half of these tasks involve sub-partitions of the misc block. We observe that as in the smaller match task blocking-based partitioning enables a linear speedup for the whole range of up to 4 nodes and 16 cores for both match strategies. We could thus reduce the execution time from approx. 6 hours to merely 24 minutes for the WAM and from 8 hours to 51 minutes for the LRM.

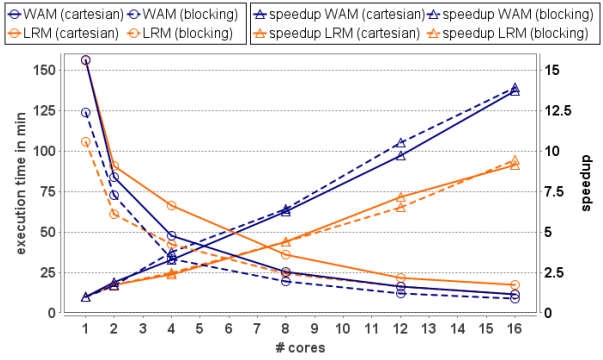
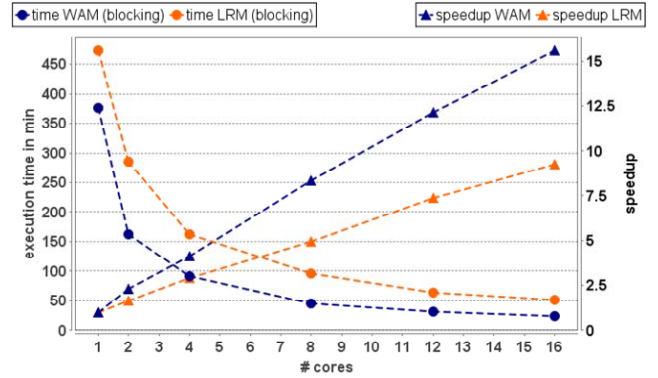

Figure 8: Speedup small scale match problem

Figure 9: Speedup large scale match

## 5.4 Parallel matching with caching

We finally study how caching of entity partitions and affinity-based scheduling of match tasks, described in Section 4, impact performance. For this experiment we focus on the large-scale match task and the use of blocking. In this case we had 306 partitions including 7 misc blocks. We used a small cache size of 16 partitions per match node (c=16), i.e. about 5% of the input dataset.

The match execution times for different number of cores and the use of caching are shown in Table 1 (WAM) and Table 2 (LRM). Each Table compares the non-caching execution time ($t_{nc}$) with the execution time using caching ($t_c$). A delta ($\Delta$) displays the difference between both; a ratio between $\Delta$ and $t_{nc}$ measures the benefit (improvement) by using the cache. The cache hit ratio *hr* shows the percentage of data partition accesses that are served from the cache (cache hits), i.e. without fetching the data from the data service.

We observe that caching and affinity-based task scheduling lead to very high hit ratios (76-83%) and significant execution time improvements for both match strategies. The hit ratios are very high despite the small cache size since many match tasks involve the small set of misc blocks and since affinity-based routing helped to schedule match tasks to the nodes with needed partitions in the cache. The execution times improved for both WAM and LRM by about 15% on average. The improvements are especially pronounced for non-parallel matching on 1 core where, without caching, the delays to fetch partitions from the data service directly increase the execution times. For more than 1 core we obtain similar execution time improvements with caching and therefore comparable speedup behavior than without caching.

| cores | 1 | 2 | 4 | 8 | 12 | 16 |
|---|---|---|---|---|---|---|
| $t_{nc}$ | 376 | 163 | 91 | 45 | 31 | 24 |
| $t_c$ | 278 | 147 | 81 | 39 | 28 | 22 |
| $\Delta = t_{nc} - t_c$ | 98 | 16 | 10 | 6 | 3 | 2 |
| $\Delta / t_{nc}$ | 26% | 10% | 11% | 12% | 9% | 10% |
| hr | 82% | 82% | 82% | 83% | 76% | 81% |

Table 1: Comparison of execution times (in minutes) for WAM using the blocking strategy (large scale problem)

| cores | 1 | 2 | 4 | 8 | 12 | 16 |
|---|---|---|---|---|---|---|
| $t_{nc}$ | 473 | 285 | 163 | 96 | 64 | 51 |
| $t_c$ | 386 | 251 | 144 | 79 | 56 | 45 |
| $\Delta = t_{nc} - t_c$ | 87 | 34 | 18 | 17 | 9 | 6 |
| $\Delta / t_{nc}$ | 18% | 12% | 11% | 17% | 13% | 12% |
| hr | 79% | 81% | 79% | 79% | 81% | 79% |

Table 2: Comparison of execution times (in minutes) for LRM using the blocking strategy (large scale problem)

## 6. RELATED WORK

Entity matching is a very active research area and many approaches have been proposed and evaluated as described in recent surveys [1, 6, 11, 14]. However, only a few approaches consider parallelized entity matching [3, 4, 10, 19]. In [4] the authors describe first ideas for parallelizing entity matching in the Febrl system. The initial implementation uses the Message Passing Interface (MPI) standard and was evaluated on a single compute node. D-Swoosh [3] does not independently match pairs of entities but iteratively merges together matching entities and uses the merged entities for further matching. This execution model is more difficult to distribute on several processors because it requires communication between match tasks to interchange merged entities. They propose different strategies for this extension for both evaluating the Cartesian product and the use of blocking. [10] proposes parallel linkage algorithms using match/merge for three input cases (clean-clean, clean-dirty, dirty-dirty) without considering blocking. Their partition strategy only splits the larger input source into multiple pieces which are distributed together with the smaller input source to available processors. The implementation is based on distributed MATLAB.

Our model of match processing is easier to distribute and usable in combination with different partitioning strategies. We consider memory requirements of different match strategies and the available computing environment for an optimal data partitioning. Unique features of our blocking-based partitioning include the proposed partition tuning and the consideration of non-partitionable entities (misc block). The D-Swoosh performance was determined only for emulated distributed environments while we use a real implementation. Furthermore, D-Swoosh and [10] evaluated smaller match tasks (5,000 – 50,000 entities) in contrast to our large match task of 114,000 entities.

Recently a first approach for parallel entity matching on a cloud infrastructure has been proposed in [19]. In particular, the authors explain how token-based similarity functions (e.g. N-Gram, PPJoin++) on a single attribute can be calculated on the popular MapReduce implementation Hadoop [8]. The approach is based on a complex workflow consisting of a preprocessing of all tokens and utilizing a token-based dynamic data redistribution for parallel matching. While the authors show the applicability of the MapReduce model for parallel entity resolution, the overall approach has to be specified at a low programming level making the resulting code hard to maintain and reuse, as already observed in [17, 18]. The token-based data redistribution can lead to large differences in partition sizes and therefore load balancing problems. Furthermore, large partitions for frequent tokens may not fit into memory causing performance degradations. The relatively low speedup values (about 4.2 for 10 cores) are further influenced by an expensive materialization of temporary results between map and reduce tasks.

By contrast we are not focusing on parallelizing specific matchers but propose a more general model that allows the parallel processing of complex match strategies that may contain several matchers. The parallel processing is based on general partitioning strategies that take memory and load balancing requirements into account thereby supporting a good speedup for parallel entity matching. Further performance benefits are achieved by caching entities and applying an affinity-based scheduling of match tasks. In contrast to [19] we consider the use of blocking strategies to reduce the search space. Transferring the proposed approach to a cloud environment is left for future work.

## 7. CONCLUSIONS AND FUTURE WORK

We propose two general partitioning strategies, size-based and blocking-based partitioning, for parallel entity matching on single or multiple input sources. These approaches can be used in combination with different mach strategies including the use of learner-based combinations of multiple matchers. Size-based partitioning is used to evaluate the Cartesian product in parallel. We choose the partition size according to the memory requirements of a match strategy. Blocking-based partitioning is used in combination with blocking to reduce the search space and applies a partition tuning for efficient parallel matching. The partitioning strategies are implemented and evaluated on a newly developed service-based infrastructure for parallel entity matching. The evaluation on large product collections shows the high effectiveness and scalability of the proposed partitioning strategies for both parallel matching within multi-core nodes and on multiple nodes. Hence, we are able to significantly reduce the execution times for different match strategies. The use of partition caches at match nodes and affinity-based scheduling of match tasks also improved performance.

In future work, we plan to adapt our approaches to cloud architectures for parallel blocking and matching. Moreover, we will investigate optimizations within match strategies, e.g., to execute different matchers in parallel.